\documentclass[10pt]{amsart}
\usepackage{graphicx,color}
\usepackage{url}

\usepackage{mathrsfs}

\newtheorem{theorem}{Theorem}[section]
\newtheorem{lemma}[theorem]{Lemma}           
\newtheorem{cor}[theorem]{Corollary}
\newtheorem{prop}[theorem]{Proposition}

\theoremstyle{definition}

\theoremstyle{remark}

\numberwithin{equation}{section}


\title[Generalized eigenfunctions for QW via path counting approach]
{Generalized eigenfunctions for quantum walks via path counting approach}
\author[T. Komatsu]{Takashi Komatsu}
\address[T. Komatsu]{Department of Bioengineering, School of Engineering,The University of Tokyo, Bunkyo-ku, Tokyo, 113-8656, Japan}
\email{komatsu@coi.t.u-tokyo.ac.jp}
\author[N. Konno]{Norio Konno}
\address[N. Konno]{Department of Applied Mathematics, Yokohama National University, Hodogaya, Yokohama, Kanagawa, 240-8501, Japan}
\email{konno-norio-bt@ynu.ac.jp}
\author[H. Morioka]{Hisashi Morioka}
\address[H. Morioka]{Graduate School of Science and Engineering,
Ehime University, Bunkyo-cho 3, Matsuyama, Ehime, 790-8577, Japan}
\email{morioka@cs.ehime-u.ac.jp}
\author[E. Segawa]{Etsuo Segawa}
\address[E. Segawa]{Graduate School of Environment Information Sciences, Yokohama National University, Hodogaya, Yokohama, Kanagawa, 240-8501, Japan}
\email{segawa-etsuo-tb@ynu.ac.jp}

\date{\today}

\begin{document}
\maketitle

\begin{abstract} 
We consider the time-independent scattering theory for time evolution operators of one-dimensional two-state quantum walks.
The scattering matrix associated with the position-dependent quantum walk naturally appears in the asymptotic behavior at spatial infinity of generalized eigenfunctions.
The asymptotic behavior of generalized eigenfunctions is a consequence of an explicit expression of the Green function associated with the free quantum walk.
When the position-dependent quantum walk is a finite rank perturbation of the free quantum walk, we derive a kind of combinatorial constructions of the scattering matrix by counting paths of quantum walkers.
We also mention some remarks on the tunneling effect.

\end{abstract}

%
%
\section{Introduction}

Discrete time quantum walks (QWs for short) have been investigated in both finite and infinite systems.
For finite systems, the effectiveness and universality of quantum walks in the quantum search algorithms have been studied (see \cite{Am}, \cite{Ch}, \cite{Po} and its references therein).
For infinite systems, there are several mathematical works obtaining limiting behaviors different from those of classical random walks (see \cite{Ko08}).
Recently, spectral and scattering theory for QWs have been intensively studied.
For example, see \cite{FeHi1}, \cite{FeHi2}, \cite{MMOS}, \cite{MS41}, \cite{MS42}, \cite{Mo}, \cite{MoSe}, \cite{RST}, \cite{Su}, \cite{Ti}.

In this paper, we consider one-dimensional position-dependent QWs in view of the scattering theory.
In our previous works \cite{Mo} and \cite{MoSe}, the method of our study was based on the scattering theory of quantum mechanics like Schr\"{o}dinger equations.
For general information of this research area, the monograph by Yafaev \cite{Ya} is available and its reference is also worthwhile.
In order to construct a generalized eigenfunction of the time evolution operator of QWs, we use the time-independent scattering theory of quantum mechanics.
We also adopt another approach in the latter half of this paper. 
Namely, we introduce a combinatorial construction of the scattering matrix associated with perturbations of finite rank.
A primitive form of this method can be seen in \cite{FeyHi}.
In our construction, we relate the generalized eigenfunction with the time evolution of the QW, and we can compute the scattering matrix by counting paths of quantum walkers.
For Schr\"{o}dinger operators with finite rank perturbations, the representation of the scattering matrix has been derived by \cite{Ku}.



Let us introduce our model of discrete time QWs.
In the following, ${\bf Z}$, ${\bf R}$ and ${\bf C}$ denote the set of integers, the set of real numbers and the set of complex numbers, respectively.
We focus on two-state QWs on ${\bf Z}$.
Let $\psi =\{ \psi (x) \} _{x\in {\bf Z}} $ be a ${\bf C}^2 $-valued sequence on ${\bf Z}$.
In the following, we denote the column vector $\psi (x)$ by $\psi (x)=[ \psi_L (x), \psi _R (x)]^{\mathsf{T}} $ for ${\bf C}$-valued sequences $\psi_L$ and $\psi_R $ on ${\bf Z}$.
Here $[a , b]^{\mathsf{T}}$ means the transpose of the row vector $[a,b]$.
First of all, we define the \textit{homogeneous QW}.
We define the operator $U_0$ by
$$
(U_0 \psi )(x)= P_0  \psi (x+1) + Q_0  \psi (x-1 ) ,
$$
where the matrices $ P_0 $ and $ Q_0 $ are given by 
\begin{gather*}
P_0 = e^{i\gamma /2} \left[ \begin{array}{cc} 
pe^{i(\alpha -\gamma /2 )} & q e^{i(\beta -\gamma /2 )} \\ 0 & 0 \end{array} \right] , \\
Q_0 = e^{i\gamma/2} \left[ \begin{array}{cc} 0 & 0 \\ -q e^{-i(\beta -\gamma/2)} & p e^{-i (\alpha -\gamma/2)} \end{array} \right] ,
\end{gather*}
for $p\in (0,1]$, $q\in [0,1)$, $p^2 +q^2 =1 $ and $\alpha , \beta , \gamma \in {\bf T} := {\bf R} /2\pi {\bf Z}$.
We have $ C_0 := P_0 + Q_0 \in \mathrm{U} (2)$.
The operator $U_0$ is unitary on $\ell^2 ({\bf Z};{\bf C}^2 )$.
Note that we can write $U_0 = SC_0 $ for the shift operator 
$$
(S\psi )(x)= \left[ \begin{array}{c} \psi_L (x+1) \\ \psi _R (x-1 ) \end{array} \right] .
$$
In Sections 3 and 4, we consider the \textit{free QW} defined by $U_0 = S$.
This is a special case of homogeneous QWs such that $C_0$ is the $2\times 2$ identity matrix.

The \textit{position-dependent QW} defined by the perturbed operator $U=SC$ is given by the operator $C$ of multiplication by the matrix $C(x)\in \mathrm{U} (2)$ for every $x\in {\bf Z} $.
Thus the operator $U$ is also unitary in $\ell^2 ({\bf Z};{\bf C}^2 )$.
The time evolutions of the homogeneous QW and the position-dependent QW are given by 
$$
\Psi (t,\cdot )=U^t \psi , \quad \Psi^{(0)} (t,\cdot)= U_0^t \psi , \quad t\in {\bf Z},
$$
for an initial state $\psi $.
Note that these time evolution preserve the norm of the initial state in $\ell^2 ({\bf Z};{\bf C}^2)$.

In this paper, we assume that the following assumptions hold : 

\medskip

{\bf (A-1)} There exist constants $\epsilon_0  , c_0 >0$ such that 
$$
\| C(x)-C_0 \| _{\infty} \leq c_0 (1+|x|)^{-1-\epsilon_0 } , \quad x\in {\bf Z} ,
$$
where $\| \cdot \| _{\infty} $ is the maximum norm for 2 by 2 matrices.

\medskip

{\bf (A-2)} For all $x\in {\bf Z} $, $C(x)$ is not anti-diagonal.

\medskip

In Section 4, the assumption (A-1) will be replaced by ``$U-U_0 $ is an operator of finite rank".

In this paper, we consider the time-independent scattering theory for $U$ and $ U_0$.
In particular, we focus on the scattering matrix associated with the wave operators
$$
W_{\pm} = {\mathop{{\rm s\text{-}lim}}_{t\to\pm \infty}} \, U^{-t} U_0^t  \quad \text{in} \quad \ell^2 ({\bf Z} ; {\bf C}^2 ) .
$$
Existence and completeness of $W_{\pm} $ for $U$ and $U_0$ have already been proven by Suzuki \cite{Su}.
Note that the assumption (A-1) is the well-known \textit{short-range condition} which guarantees the existence and the asymptotic completeness of the wave operators.

\begin{theorem}[Suzuki \cite{Su}]
The wave operators exist and are complete i.e. the ranges of $W_{\pm} $ coincide with the absolutely continuous subspace $\mathcal{H}_{ac} (U)$ of $\ell^2 ({\bf Z};{\bf C}^2 )$ for $U$.
Precisely, for any $\phi \in \mathcal{H}_{ac} (U)$, there exist $\psi _{\pm} \in \ell^2 ({\bf Z};{\bf C}^2 )$ such that $\| U_0^t \psi _{\pm} - U^t \phi \| _{\ell^2 ( {\bf Z};{\bf C}^2 )} \to 0$ as $t\to \pm \infty $.
The wave operators are unitary on $\ell^2 ({\bf Z};{\bf C}^2 )$ and we have $W_{\pm}^* = W_{\pm}^{-1} $. 

\label{S1_thm_suzuki}
\end{theorem}

The assumption (A-2) guarantees the penetrability of barriers given by $C(x)$.
In fact, if $C(x)$ is anti-diagonal at a point $x=x_0$, a quantum walker is reflected at $x_0$.

The scattering operator is defined by 
$$
\Sigma = W_+^* W_- : \psi _- \mapsto \psi_+ .
$$
Its Fourier transform $\widehat{\Sigma} $ is decomposed by the scattering matrix :
$$
\widehat{\Sigma} = \int _{J_{\gamma}} \oplus \widehat{\Sigma} (\theta ) d\theta ,
$$
where $ J_{\gamma} $ is the interval defined in Lemma \ref{S2_lem_essspec} (see also \cite{RST} or \cite{Mo}).
The scattering matrix (S-matrix for short) $\widehat{\Sigma} (\theta )$ naturally appears in the generalized eigenfunction for $U$.
One of the purposes of this paper is to derive the asymptotic behavior as $x \to \pm \infty $ of the generalized eigenfunction in $\ell^{\infty} ({\bf Z};{\bf C}^2 )$.
The S-matrix can be represented by the distorted Fourier transformation associated with $U$ and $U_0$.
The distorted Fourier transformation is constructed as the spectral decomposition of unitary operators.

We also mention stationary measures of QWs as a related topic.
Generalized eigenfunctions of $U$ give stationary measures for $U$.
Note that this is a special kind of stationary measures.
For the topic of stationary measures, see e.g. Konno \cite{Ko08}, Konno-Takei \cite{KoTa}, Komatsu-Konno \cite{KoKo}, Kawai et al. \cite{K3}.

The plan of this paper is as follows.
In Section 2, we consider the Green function of the homogeneous QW.
Green functions have been derived in the momentum space by \cite{Mo}.
In this paper, we derive an explicit formula of Green functions on ${\bf Z}$.
In Section 3, we construct the generalized eigenfunction for $U$ in $\ell^{\infty} ({\bf Z};{\bf C}^2 )$. 
The optimal functional space in which there exist generalized eigenfunctions is characterized by Agmon-H\"{o}rmander's $\mathcal{B}$-$\mathcal{B}^*$ spaces (\cite{AgHo}).
By using the formula of the Green function, we obtain the asymptotic behavior at infinity of generalized eigenfunctions.
In Section 4, instead of general theory of the time-independent scattering theory, we introduce a combinatorial construction of the S-matrix when $U-U_0$ is an operator of finite rank.
The scattering matrix is computed by a finite rank submatrix induced from the operator $U$.
We also mention the resonant-tunneling effect (see also \cite{MMOS}) here.
Finally, we prove that incident waves always pass through penetrable barriers.
Some remarks on complex contour integrals are gathered in Appendix A.

\subsection{Notation}

The notation in this paper is as follows.
${\bf T} := {\bf R}/2\pi {\bf Z} $ denotes the flat torus.
Similarly, the complex flat torus is denoted by ${\bf T} _{{\bf C}} = {\bf C}/2\pi {\bf Z} = \{ x+iy \ ; \ x\in {\bf T} , y\in {\bf R} \} $.
We often use the identifications ${\bf T} = [-\pi , \pi )$ or $[0,2\pi )$ under $2\pi$-periodicity.
$ {\bf T} _{{\bf C}}$ appears in Appendix A.
For a vector $ {\bf v} \in {\bf C}^2 $, $|{\bf v}|$ denotes the norm in ${\bf C}^2 $. 
For two vectors $ {\bf v}, {\bf w}\in {\bf C}^2 $, we denote by $\langle {\bf v} , {\bf w} \rangle $ the inner product in ${\bf C}^2 $.
For a ${\bf C}^2$-valued sequence $f=\{ f(x) \} _{x\in {\bf Z}} $, we define the mapping $\mathcal{U}$ by the Fourier transformation
$$
\widehat{f} (\xi ):= (\mathcal{U}f )(\xi )= \frac{1}{\sqrt{2\pi}} \sum _{x\in {\bf Z}} e^{-ix\xi} f(x), \quad \xi \in {\bf T}.
$$
The Fourier coefficients of $\widehat{g} (\xi )$ on ${\bf T}$ is given by 
$$
g(x)= (\mathcal{U}^* \widehat{g} )(x)=\frac{1}{\sqrt{2\pi}} \int _{{\bf T}} e^{ix\xi} \widehat{g} (\xi )d\xi , \quad x\in {\bf Z} .
$$
Then $\mathcal{U}$ is a unitary transformation from $\ell^2 ({\bf Z} ; {\bf C}^2 )$ to $L^2 ({\bf T}; {\bf C}^2 )$.
Here $\ell^2 ({\bf Z} ; {\bf C}^2 )$ and $L^2 ({\bf T}; {\bf C}^2 )$ are equipped with its inner products and norms
$$
(f,g) _{\ell^2 ({\bf Z};{\bf C}^2 )} = \sum _{x\in {\bf Z}} \langle f(x),g(x)\rangle , \quad \| f\|_{\ell^2 ({\bf Z};{\bf C}^2 )} =\sqrt{ (f,f) _{\ell^2 ({\bf Z};{\bf C}^2 )} },
$$
$$
(\widehat{f},\widehat{g} ) _{L^2 ({\bf T} ;{\bf C}^2 )} = \int _{{\bf T}} \langle \widehat{f} (\xi ),\widehat{g} (\xi )\rangle d\xi , \quad \| \widehat{f} \|_{L^2 ({\bf T} ; {\bf C}^2 )} = \sqrt{(\widehat{f},\widehat{f} ) _{L^2 ({\bf T} ;{\bf C}^2 )}} .
$$
For Banach spaces $X$ and $ Y$, ${\bf B} (X;Y)$ denotes the space of bounded linear operators from $X$ to $Y$.


\section{Green function}

\subsection{Resolvent for the homogeneous QW}
The spectral theory for $U$ and $U_0$ is studied in \cite{RST}, \cite{Mo}, \cite{MoSe}, \cite{MS42}.
Here let us list some basic results.

Let
$$
\widehat{U}_0 = \mathcal{U} U_0 \mathcal{U}^* .
$$
It follows that $\widehat{U}_0$ is the operator of multiplication on ${\bf T}$ by the unitary matrix
$$
\widehat{U}_0 (\xi )= e^{i\gamma/2} \left[ \begin{array}{cc} pe^{i(\alpha -\gamma/2 )} e^{i\xi} & qe^{i(\beta -\gamma/2 )} e^{i\xi} \\ -q e^{-i(\beta -\gamma/2)} e^{-i\xi} & pe^{-i(\alpha -\gamma/2)} e^{-i\xi} \end{array} \right] , \quad \xi \in {\bf T} . 
$$ 
For any $\kappa \in {\bf C} $, we have 
\begin{gather*}
\begin{split}
p(\xi ,\kappa ):=& \, \mathrm{det} (\widehat{U}_0 (\xi )-e^{i\kappa} ) \\ 
=& \,  2pe^{i(\kappa +\gamma/2)} \left( -\cos \left( \xi + \alpha -\frac{\gamma}{2}  \right) +\frac{1}{p} \cos \left( \kappa-\frac{\gamma}{2}  \right) \right) .
\end{split}
\end{gather*}
By using $p(\xi ,\kappa )$, we can evaluate the continuous spectrum of $ U_0 $.
Moreover, Weyl's singular sequence method shows the essential spectrum of $U$.
For a unitary operator $A$, let $\sigma (A)$ denote the spectrum of $A$. 
We denote by $\sigma_{ess} (A)$, $\sigma_d (A)$ the essential spectrum and the discrete spectrum of $A$.
By using the spectral decomposition of $A$, we can also define the absolutely continuous spectrum $\sigma_{ac} (A)$, the point spectrum $\sigma_p (A)$, and the singular continuous spectrum $ \sigma_{sc} (A)$.
See \cite{RST}, \cite{Mo}, or \cite{MoSe} for more details of definitions.
Note that the absence of singular continuous spectrum was first proved by Asch et al. \cite{ABJ}.

\begin{lemma}
Following assertions hold.
\begin{enumerate}
\item Let $J_{\gamma}  = J_{\gamma ,1} \cup J_{\gamma ,2} $ where
\begin{gather*}
\begin{split}
&J_{\gamma,1} = [\arccos p +\gamma/2, \pi -\arccos p +\gamma /2], \\
&J_{\gamma,2} = [\pi + \arccos p +\gamma/2, 2\pi -\arccos p +\gamma /2].
\end{split}
\end{gather*}
Then we have $\sigma (U_0)= \sigma_{ac} (U_0 )= \{ e^{i\theta} \ ; \ \theta \in J_{\gamma} \} $.
\item We have $\sigma_{ess} (U)= \sigma_{ess} (U_0) =\{ e^{i\theta} \ ; \ \theta \in J_{\gamma} \} $ and $\sigma_{sc} (U)=\emptyset $.
\item Let $\mathcal{T} = \{ e^{i\theta} \ ; \ \theta \in J_{\gamma,\mathcal{T}} \} $ where 
\begin{gather*}
\begin{split}
J_{\gamma,\mathcal{T}} = \left\{
\begin{split}
&\arccos p +\gamma/2, \ \pi -\arccos p +\gamma /2 , \\
&\pi + \arccos p +\gamma/2, \ 2\pi -\arccos p +\gamma /2
\end{split}
\right\} .
\end{split}
\end{gather*}
Then there is no eigenvalue of $U$ in $\sigma_{ess} (U)\setminus \mathcal{T} $.
\item For $e^{i\theta} \in \mathcal{T}$, the zeros of $p(\xi ,\theta )$ are degenerate i.e. $\frac{\partial p}{\partial \xi} (\xi , \theta )=0$ if $p(\xi ,\theta )=0$.
For $e^{i\theta} \in \sigma(U_0)\setminus \mathcal{T}$, the zeros of $p(\xi ,\theta )$ are non-degenerate i.e. $\frac{\partial p}{\partial \xi} (\xi , \theta ) \not= 0$ if $p(\xi ,\theta )=0$.
\end{enumerate}
\label{S2_lem_essspec}
\end{lemma}

Let us turn to the resolvent $R_0 (\kappa )=(U_0 -e^{i\kappa } )^{-1} $ for $\kappa \in {\bf C}\setminus {\bf R}$.
The operator $\widehat{R}_0 (\kappa )=\mathcal{U} R_0 (\kappa ) \mathcal{U}^* $ is the operator of multiplication on ${\bf T}$ by the matrix
\begin{equation}
\widehat{R}_0 (\xi ,\kappa )=\frac{1}{p(\xi ,\kappa )} \left[ \begin{array}{cc} pe^{-i(\alpha -\gamma )}e^{-i\xi} -e^{i\kappa } & -qe^{i\beta}e^{i\xi} \\ qe^{-i(\beta -\gamma )} e^{-i\xi} & pe^{i\alpha}e^{i\xi} -e^{i\kappa } \end{array} \right] .
\label{S2_eq_resolventmatrixxi}
\end{equation}
In \cite{Mo}, we have proven the limiting absorption principle of $R_0 (\kappa )$ as follows.
Agmon-H\"{o}rmander's $\mathcal{B}$-$\mathcal{B}^*$ spaces (\cite{AgHo}) are often used in the context of the limiting absorption principle.
The functional spaces $\mathcal{B} ({\bf Z})$ and $\mathcal{B}^* ({\bf Z})$ of ${\bf C}^2$-valued sequences on ${\bf Z}$ are defined by the norms
$$
\| f\| _{\mathcal{B} ({\bf Z} )} = \sum _{j=0}^{\infty} r_j^{1/2} \left( \sum _{r_{j-1} \leq |x|< r_j} |f(x)|^2  \right)^{1/2} ,
$$
for $r_{-1}=0$, $r_j =2^j$, $j\geq 0$, and
$$
\| u\|_{\mathcal{B}^* ({\bf Z})}^2 = \sup _{R>1} \frac{1}{R} \sum _{|x|<R} |u(x)|^2 .
$$
Note that the inclusion relation
$$
\mathcal{B} ({\bf Z} )\subset \ell^2 ({\bf Z};{\bf C}^2)\subset \mathcal{B}^* ({\bf Z} )
$$
holds.

\begin{lemma}
Following assertions hold.
\begin{enumerate}
\item For $ \theta \in J_{\gamma} \setminus J_{\gamma,\mathcal{T}} $ and $f\in \mathcal{B} ({\bf Z} )$, there exist weak-$*$ limits $R_0 (\theta \pm i0 )f := \lim _{\epsilon \downarrow 0 } R_0 ( \theta -i\log (1\mp \epsilon ))f$ in the sense
$$
\lim _{\epsilon \downarrow 0} (R_0 ( \theta -i\log (1\mp \epsilon ))f,g) = (R_0 (\theta \pm i0)f,g), \quad g\in \mathcal{B} ({\bf Z} ).
$$

\item $R_0 ( \theta \pm i0 )f$ for $f\in \mathcal{B} ({\bf Z} )$ satisfy $\| R_0 (\theta \pm i0 )f\| _{\mathcal{B}^* ({\bf Z} ) } \leq c\| f\| _{\mathcal{B} ({\bf Z} )} $ for a constant $c>0$ which is independent of $\theta$ if $\theta$ varies over a compact interval in $J_{\gamma} \setminus J_{\gamma,\mathcal{T}} $.
\item The mappings $J_{\gamma} \setminus J_{\gamma,\mathcal{T}} \ni \theta \mapsto (R_0 (\theta \pm i0 )f,g) $ are continuous for $f,g\in \mathcal{B} ({\bf Z} )$.
\end{enumerate}
\label{S2_lem_LAP0}
\end{lemma}

\subsection{Green function}

In \cite{Mo}, we also derived the radiation condition which guarantees the uniqueness of the solution to the equation
$$
(U_0 -e^{i\theta} )u=f \quad \text{on} \quad {\bf Z}, \quad \theta \in J_{\gamma} \setminus J_{\gamma,\mathcal{T}} ,
$$
in $\mathcal{B}^* ({\bf Z} )$ for $f\in \mathcal{B} ({\bf Z} )$.
However, our result was proven in the momentum space i.e. $\xi$-space.
In the following, we consider this equation in the physical space i.e. $x$-space.

For $\kappa \in {\bf C}\setminus {\bf R} $, we consider the equation 
\begin{equation}
(U_0 -e^{i\kappa} )u=f \quad \text{on} \quad {\bf Z}, 
\label{S2_eq_helmholtz0}
\end{equation}
for $f\in \mathcal{B} ({\bf Z} )$.
Since $e^{i\kappa} \not\in \sigma (U_0)$, the resolvent $ R_0 (\kappa ) $ exists in $\ell^2 ({\bf Z};{\bf C}^2 )$.
We seek the solution to the equation (\ref{S2_eq_helmholtz0}) of the form 
\begin{gather}
u (x)= \sum _{y\in {\bf Z}} G_0 (x-y,\kappa ) f(y) ,
\label{S2_eq_helmholtz0sol}
\end{gather}
where the kernel $G_0 (x,\kappa )$ is the matrix
$$
G_0 (x,\kappa ) = \left[ \begin{array}{cc} r_{11} (x,\kappa ) & r_{12} (x,\kappa ) \\ r_{21} (x,\kappa ) & r_{22} (x,\kappa ) \end{array} \right] .
$$

The following lemma is a consequence of the representation (\ref{S2_eq_helmholtz0sol}).

\begin{lemma}
Let $\delta^{(L)} =[\delta ,0]^{\mathsf{T}}$ and $\delta^{(R)}  =[0,\delta]^{\mathsf{T}}$ for $\delta = \{ \delta_{x0} \} _{x\in {\bf Z} } $ where $\delta _{x0} $ is the Kronecker delta.
The solutions $u^{(L)}$ and $u^{(R)}$ to the equations
\begin{equation}
(U_0 -e^{i\kappa} )u^{(L)} = \delta^{(L)} ,\quad (U_0 -e^{i\kappa} )u^{(R)}= \delta ^{(R)} ,
\label{S2_eq_S2_eq_helmholtz011}
\end{equation}
are given by 
\begin{equation}
u^{(L)} (x)= \left[ \begin{array}{c} r_{11} (x,\kappa ) \\ r_{21} (x,\kappa ) \end{array} \right] , \quad u^{(R)} (x)=  \left[ \begin{array}{c} r_{12} (x,\kappa ) \\ r_{22} (x,\kappa ) \end{array} \right] .
\label{S2_eq_S2_eq_helmholtz012}
\end{equation}
\label{S2_lem_kernel}
\end{lemma}

Now let us compute an explicit expression of $G_0 (x,\kappa )$ by using Lemma \ref{S2_lem_kernel}.
In view of (\ref{S2_eq_S2_eq_helmholtz011}), we have 
$$
(\widehat{U}_0 (\xi )-e^{i\kappa} ) \widehat{u}^{(L)} (\xi )=\frac{1}{\sqrt{2\pi}} \left[ \begin{array}{c} 1 \\ 0 \end{array} \right] , \quad (\widehat{U}_0 (\xi )-e^{i\kappa} ) \widehat{u}^{(R)} (\xi )=\frac{1}{\sqrt{2\pi}} \left[ \begin{array}{c} 0 \\ 1 \end{array} \right] .
$$
Due to (\ref{S2_eq_resolventmatrixxi}), we have 
\begin{gather}
r_{11}  (x,\kappa )=\frac{1}{2\pi} \int _{{\bf T}} \frac{e^{ix\xi} (pe^{-i(\xi+\alpha -\gamma )}-e^{i\kappa})}{p(\xi,\kappa )} d\xi , \label{S2_lem_kernelr11} \\
r_{21} (x,\kappa )=\frac{1}{2\pi} \int _{{\bf T}} \frac{qe^{ix\xi} e^{-i(\xi+\beta-\gamma )}}{p(\xi,\kappa )} d\xi , \label{S2_lem_kernelr21}
\end{gather}
and
\begin{gather}
r_{12}  (x,\kappa )=-\frac{1}{2\pi} \int _{{\bf T}} \frac{qe^{ix\xi} e^{i(\xi+\beta )} }{p(\xi,\kappa )} d\xi , \label{S2_lem_kernelr12} \\
r_{22} (x,\kappa )=\frac{1}{2\pi} \int _{{\bf T}} \frac{e^{ix\xi} (pe^{i(\xi+\alpha)} -e^{i\kappa} )}{p(\xi,\kappa )} d\xi . \label{S2_lem_kernelr22}
\end{gather}

We compute the integrals (\ref{S2_lem_kernelr11})-(\ref{S2_lem_kernelr22}) by using the result in Appendix A.
In fact, we have
\begin{gather}
r_{11}  (x,\kappa )=\frac{1}{2\pi} \frac{e^{-ix(\alpha -\gamma /2)}}{2pe^{i(\kappa +\gamma/2)}} \left( pe^{i\gamma/2} I(x-1, \kappa ) -e^{i\kappa} I(x,\kappa ) \right) ,
 \label{S2_lem_kernelr11*} \\
r_{21} (x,\kappa )=\frac{1}{2\pi} \frac{qe^{-i(\beta -\gamma )}}{2pe^{i(\kappa +\gamma /2)}} I (x-1 ,\kappa ), \label{S2_lem_kernelr21*}
\end{gather}
and
\begin{gather}
r_{12}  (x,\kappa )=-\frac{1}{2\pi}  \frac{qe^{i\beta}e^{i(x+1)(-\alpha + \gamma /2)}}{2pe^{i(\kappa +\gamma/2)}} I(x+1,\kappa ), \label{S2_lem_kernelr12*} \\
r_{22} (x,\kappa )=\frac{1}{2\pi}  \frac{e^{-ix(\alpha -\gamma /2)}}{2pe^{i(\kappa +\gamma/2)}} \left( pe^{i\gamma/2} I(x+1,\kappa ) -e^{i\kappa} I(x,\kappa )  \right) , \label{S2_lem_kernelr22*}
\end{gather}
where $I(x,\kappa )$ denotes the integral (\ref{App_integral}).
In view of Lemma \ref{App_lem_contourintegral}, we can derive the formula of $ G_0 (x,\kappa )$.

\begin{lemma}
For $\kappa \in {\bf C}\setminus {\bf R}$, we have
\begin{gather}
r_{11} (x,\kappa )=\pm \frac{ie^{ix(-\alpha +\gamma/2) } (p e^{\pm i |x-1|\zeta (\kappa )} -e^{i(\kappa -\gamma/2)} e^{\pm i |x|\zeta (\kappa )})}{2pe^{i\kappa } \sin \zeta (\kappa )} , \label{S2_eq_Green11} \\
r_{21} (x,\kappa )= \pm \frac{iq e^{i(\alpha -\beta)}  e^{ix(- \alpha +\gamma/2)}   e^{\pm i |x-1| \zeta (\kappa )}}{2pe^{i\kappa } \sin \zeta (\kappa )} ,\label{S2_eq_Green21}
\end{gather}
and
\begin{gather}
r_{12} (x,\kappa )= \mp \frac{iq  e^{i(-\alpha + \beta)}  e^{ix(-\alpha +\gamma/2)}   e^{\pm i |x+1| \zeta (\kappa )}}{2pe^{i\kappa } \sin \zeta (\kappa )} , \label{S2_eq_Green12} \\
r_{22} (x,\kappa )=\pm \frac{ie^{ix( -\alpha + \gamma/2)} (p e^{\pm i |x+1|\zeta (\kappa)} -e^{i(\kappa -\gamma/2)} e^{\pm i |x| \zeta (\kappa )})}{2pe^{i\kappa } \sin \zeta (\kappa )} ,\label{S2_eq_Green22}
\end{gather}
for $\pm \mathrm{Im} \, \zeta (\kappa )>0 $ respectively where $\zeta (\kappa )$ is defined by (\ref{S2_eq_MCK}).
\label{S2_lem_formula_Green}
\end{lemma}

The formula of $ G_0 (x,\kappa )$ is not so simple.
In the following, we consider the case where $ U_0 $ is the free QW for the sake of simplicity. 
In this case, we have $J_{0,1} = [0,\pi ]$, $J_{0,2} =[\pi,2\pi ]$, $J_{0,\mathcal{T}} = \{0,\pi,2\pi \}$.
We can take $\zeta (\kappa )= \kappa $ for $ 0< \mathrm{Re} \, \kappa < \pi$ and $\zeta (\kappa )=2\pi -\kappa$ for $\pi <\mathrm{Re} \, \kappa <2\pi $.
Taking $\kappa =\kappa_{\pm}= \theta -i \log (1\mp \epsilon )$ for $\theta \in J_{0} \setminus J_{0,\mathcal{T}}$ and $\epsilon >0$, we have 
\begin{gather*}
r_{11} (x,\kappa _{\pm} )=\pm \frac{i(e^{\pm i \kappa_{\pm} |x-1|} -e^{\pm i \kappa_{\pm} (|x|\pm 1)})}{2e^{i\kappa_{\pm} } \sin \kappa_{\pm} } ,\\
r_{22} (x,\kappa _{\pm} )=\pm \frac{i(e^{\pm i \kappa_{\pm}
 |x+1|} -e^{\pm i \kappa _{\pm} (|x|\pm 1)})}{2e^{i\kappa_{\pm}} \sin \kappa_{\pm} } ,\\
r_{12 } (x,\kappa _{\pm} )=r_{21} (x,\kappa_{\pm} )=0.
\end{gather*}
More precisely, we obtain the following explicit expression of $G_0 (x,\kappa _{\pm} )$.

\begin{lemma}

Suppose $ U_0 =S $.
Let $F(x)=1 $ for $x\geq 0$ and $F(x)=0$ for $x\leq -1$.
The corresponding diagonal kernel $G_0 (x,\kappa _{\pm} )$ is given by 
\begin{gather}
r_{11} (x,\kappa_+ )= F(x-1) e^{i\kappa_+ (x-1)} , \quad  r_{22} (x,\kappa_+ )= F(-x-1) e^{-i\kappa_+ (x+1)} , \label{S2_eq_freeGreen+1} \\
r_{11} (x,\kappa_- )= -F(-x)e^{i\kappa_+ (x-1)} , \quad r_{22} (x,\kappa_- )=- F(x)e^{-i\kappa_- (x+1)} ,
\label{S2_eq_freeGreen-1}
\end{gather}
for $ \theta \in J_{0} \setminus J_{0, \mathcal{T}} $.
\label{S2_lem_freeGreen}
\end{lemma}

\textit{Remark.}
We can obtain Lemma \ref{S2_lem_freeGreen} directly from 
$$
r_{11}  (x,\kappa )=\frac{1}{2\pi} \int _{-\pi}^{\pi} \frac{e^{ix\xi}}{e^{i\xi} -e^{i\kappa}}d\xi , \quad r_{22} (x,\kappa )=\frac{1}{2\pi} \int _{-\pi}^{\pi} \frac{e^{ix\xi}}{e^{-i\xi} -e^{i\kappa}} d\xi ,
$$
by using the complex contour integral.

\medskip

The formulas (\ref{S2_eq_freeGreen+1})-(\ref{S2_eq_freeGreen-1}) hold for $G_0 (x,\theta \pm i0)$, $\theta \in ( J_{0,1} \cup J_{0,2} )\setminus J_{0,\mathcal{T}} $.
We obtain the formulas
\begin{gather}
\begin{split}
&(R_0 ( \theta +i0 )f)(x) \\
&= 
e^{i\theta (x-1)} \sum _{y\leq x-1} \left[ \begin{array}{c} e^{-i\theta y} f_L (y) \\ 0 \end{array} \right] +e^{-i\theta (x+1)} \sum _{y\geq x+1} \left[ \begin{array}{c} 0 \\ e^{i\theta y} f_R (y) \end{array} \right] ,
\end{split}
\label{S2_eq_resolventfree1+}
\end{gather}
\begin{gather}
\begin{split}
&(R_0 ( \theta -i0 )f)(x) \\
&= -e^{i\theta (x-1)} \sum _{y\geq x} \left[ \begin{array}{c} e^{-i\theta y} f_L (y) \\ 0 \end{array} \right] -e^{-i\theta (x+1)} \sum _{y\leq x} \left[ \begin{array}{c} 0 \\ e^{i\theta y} f_R (y) \end{array} \right] ,
\end{split}
\label{S2_eq_resolventfree1-}
\end{gather}
for $ \theta \in J_{0} \setminus J_{0,\mathcal{T}} $ and $f\in \mathcal{B} ({\bf Z} )$.

\medskip

\textit{Remark.}
Maeda et al. \cite{MS42} have given an essentially equivalent expression of the Green function $G_0 (x,\kappa ) $ by using a computation of the transfer matrix associated with QWs (see Proposition 3.6 and Lemma 4.9 in \cite{MS42}).
Compared to our construction by the Fourier transforms, their argument gives more clearly a discrete and unitary analogue of the scattering theory for the Strum-Liouville differential equation.
On the other hand, our method can be extended to multi-dimensional QWs.


\section{Generalized eigenfunction}

\subsection{Spectral representation for Free QW}

In the following, we assume $U_0 = S$.
In \cite{Mo}, we derived a characterization of the generalized eigenfunction 
\begin{equation}
(U_0 -e^{i\theta} )u=0 \quad \text{on} \quad {\bf Z} , \quad \theta \in J_{0} \setminus J_{0,\mathcal{T}} ,
\label{S3_eq_freeef}
\end{equation}
in $\mathcal{B}^* ({\bf Z} )$.
In order to apply this theory, we recall the framework of the spectral representation for $ U_0 $.

Now we introduce the function $ \theta (\xi )= \arccos (\cos \xi )$ for $\xi \in {\bf T}$.
We have $ J_{0,1} =\{ \theta (\xi ) \ ; \ \xi \in {\bf T} \} $ and $ J_{0,2} =\{ 2\pi -\theta (\xi ) \ ; \ \xi \in {\bf T}\} $.
The matrix $\widehat{U}_0 (\xi )$ has the spectral decomposition
$$
\widehat{U}_0 (\xi )= e^{i\theta (\xi )} \widehat{ P}_1 (\xi ) + e^{-i\theta  (\xi )} \widehat{ P}_2 (\xi ) ,
$$
where 
\begin{gather*}
\widehat{P}_1 (\xi )= \left\{
\begin{split}
\left[ \begin{array}{cc} 1 & 0 \\ 0 & 0 \end{array} \right] ,& \quad 0< \xi <\pi , \\
\left[ \begin{array}{cc} 0 & 0 \\ 0 & 1 \end{array} \right] ,& \quad \pi < \xi <2\pi , 
\end{split}
\right. \quad 
\widehat{P}_2 (\xi )= \left\{
\begin{split}
\left[ \begin{array}{cc} 0 & 0 \\ 0 & 1 \end{array} \right] ,& \quad 0< \xi < \pi , \\
\left[ \begin{array}{cc} 1 & 0 \\ 0 & 0 \end{array} \right] ,& \quad \pi< \xi < 2\pi .
\end{split}
\right. 
\end{gather*}
Note that $e^{\pm i\theta (\xi )}=e^{\pm i\xi} $ for $0\leq \xi \leq \pi$ and $e^{\pm i\theta (\xi )} = e^{ \mp i\xi} $ for $\pi \leq \xi < 2\pi $.
Letting 
\begin{gather*} 
M(\theta )=\{ \xi \in {\bf T} \ ; \ p(\xi , \theta )=0\} =\{ \pm \theta (\xi ) \} ,
\end{gather*}
 we introduce the operator $\mathcal{F}_0 (\theta )= (\mathcal{F}_{0,1} (\theta), \mathcal{F}_{0,2} (\theta ))$ where 
\begin{gather*}
\mathcal{F}_{0,j} (\theta )f=\left\{
\begin{split}
\widehat{P}_j   \widehat{f} \big| _{M(\theta )} ,& \quad \theta \in J_{0,j} ,\\
0 , & \quad  \theta \not\in J_{0,j} ,
\end{split}
\right. \quad f\in \mathcal{B} ({\bf Z} ).
\end{gather*}
In order to characterize the range of $\mathcal{F}_0 (\theta )$, we introduce the vector space ${\bf h} (\theta )$ as follows.
Let $\widetilde{{\bf h}} (\theta )$ be the space of ${\bf C} $-valued functions on $M(\theta )$ with its inner product
$$
(\phi , \psi ) _{\widetilde{{\bf h} }(\theta )} = \sum _{\xi (\theta )\in M(\theta )}  \phi (\xi (\theta) ) \overline{\psi (\xi (\theta ))} ,\quad \phi , \psi \in \widetilde{{\bf h}} (\theta ).
$$
The eigenvectors ${\bf e}_1 (\xi ), {\bf e}_2 (\xi )\in {\bf C}^2$ of $\widehat{U}_0 (\xi )$ with respect to eigenvalues $e^{i\theta (\xi )} $, $e^{-i\theta (\xi )} $ are given by
\begin{gather*}
{\bf e}_1 (\xi )= \left\{
\begin{split}
\left[ \begin{array}{c} 1 \\ 0 \end{array} \right] ,& \quad 0<\xi<\pi , \\
\left[ \begin{array}{c} 0 \\ 1 \end{array} \right] ,& \quad \pi<\xi<2\pi ,
\end{split}
\right. , \quad 
{\bf e}_2 (\xi )= \left\{
\begin{split}
\left[ \begin{array}{c} 0 \\ 1 \end{array} \right] ,& \quad 0<\xi<\pi , \\
\left[ \begin{array}{c} 1 \\ 0 \end{array} \right] ,& \quad \pi<\xi<2\pi .
\end{split}
\right. 
\end{gather*}
We introduce the vector space 
\begin{gather*}
\widetilde{{\bf h}} (\theta ) {\bf e}_j = \left\{
\begin{split}
 \{ \phi {\bf e}_j \big| _{M(\theta )} \ ; \ \phi \in \widetilde{{\bf h} } (\theta ) \}  &, \quad \theta \in J_{0,j}, \\
\{ 0 \} &, \quad \theta \not\in J_{0,j},
\end{split}
\right.
\end{gather*}
and we define
$$
{\bf h} ( \theta )= \widetilde{{\bf h}} (\theta ) {\bf e}_1 \oplus  \widetilde{{\bf h}} (\theta ) {\bf e}_2 , \quad \theta \in J_0 \setminus J_{0,\mathcal{T}} .
$$
For $ \phi \in {\bf h} (\theta )$, we can take $\phi_1 , \phi_2 \in \widetilde{{\bf h}} (\theta )$ such that $\phi $ is formally represented by $\phi = \phi_1 {\bf e}_1 \oplus \phi_2 {\bf e}_2 $ where $\phi_j {\bf e}_j  \in \widetilde{{\bf h}} (\theta ) {\bf e}_j $.
Then we have $\mathcal{F}_0 (\theta ) \in {\bf B} (\mathcal{B} ({\bf Z} );{\bf h} (\theta ))$ for $ \theta \in J_{0} \setminus J_{0,\mathcal{T}} $, and we can characterize the generalized eigenfunctions of $U_0$ in $\mathcal{B}^* ({\bf Z} )$ by using the adjoint operator $\mathcal{F}_0 (\theta )^*$ as follows.

\begin{lemma}
Let $\theta \in J_0 \setminus J_{0,\mathcal{T}} $.
Then we have $\mathcal{F}_0 (\theta )^* {\bf h} (\theta )=\{ u\in \mathcal{B}^* ({\bf Z} ) \ ; \ (U_0 -e^{i\theta} )u=0\} $.
\label{S2_lem_characterizationfree}
\end{lemma}

Proof.
See Theorem 3.11 in \cite{Mo}.
\qed

\medskip

In fact, $\mathcal{F}_0 (\theta )^* \phi \in \mathcal{B}^* ({\bf Z} ) $ for the free QW is given by 
\begin{gather}
(\mathcal{F}_0 (\theta )^* \phi )(x) = 
 \frac{1}{\sqrt{2\pi}}      \left[ \begin{array}{c}  \phi_j (\theta ) e^{i\theta x}    \\ \phi_j (-\theta ) e^{-i\theta x} \end{array} \right]   , \quad \theta \in J_{0,j} \setminus J_{0,\mathcal{T}} .
\label{S3_eq_generalef000}
\end{gather}
Here we adopt the identification 
\begin{gather*}
M(\theta )\ni \pm \theta (\xi )= \left\{ 
\begin{split}
\pm \theta &, \quad \theta \in J_{0,1} \setminus J_{0,\mathcal{T}} , \\
\mp \theta &, \quad \theta \in J_{0,2} \setminus J_{0,\mathcal{T}} .
\end{split}
\right.
\end{gather*}

The operator $\mathcal{F}_0 (\theta )$ appears in the asymptotic behavior of $ R_0 (\theta \pm i0 )$ at $x\to \pm \infty$.
Here we define the equivalence relation $u\simeq v$ as $x\to \pm \infty$ for $u,v \in \mathcal{B}^* ({\bf Z} )$ by
$$
u\simeq v \Leftrightarrow u(x)-v(x)=o(1) \quad \text{as} \quad x\to \pm \infty .
$$

\begin{lemma}
Let $f\in \mathcal{B} ({\bf Z} )$.
We have 
\begin{gather}
(R_0 (\theta +i0)f)(x) \simeq \sqrt{2\pi} e^{-i\theta} e^{\pm i \theta x} (\mathcal{F}_{0,1} (\theta )f)( \pm \theta ) \quad x\to \pm  \infty , \label{S3_eq_freeasymptotic1++}   \\
( R_0 (\theta - i0 )f)(x) \simeq - \sqrt{2\pi} e^{-i\theta} e^{\mp i \theta x} (\mathcal{F}_{0,1} (\theta )f) (\mp \theta ), \quad x\to  \pm \infty , \label{S3_eq_freeasymptotic1+-}   
\end{gather}
for $ \theta \in J_{0,1} \setminus J_{0,\mathcal{T}}$, and
\begin{gather}
( R_0 (\theta +i0 )f )(x)\simeq  \sqrt{2\pi} e^{-i\theta} e^{ \pm i \theta x} (\mathcal{F}_{0,2} (\theta )f) (\pm \theta ) , \quad x\to \pm \infty , \label{S3_eq_freeasymptotic2++}  \\
 ( R_0 (\theta - i0 )f )(x)\simeq - \sqrt{2\pi} e^{-i\theta} e^{\mp i \theta x} (\mathcal{F}_{0,2} (\theta )f) (\mp \theta ) , \quad x\to \pm \infty , \label{S3_eq_freeasymptotic2+-} 
\end{gather}
for $ \theta \in J_{0,2} \setminus J_{0,\mathcal{T}} $.
\label{S3_lem_asymptopticfree}
\end{lemma}

Proof.
This lemma is a direct consequence of (\ref{S2_eq_resolventfree1+})-(\ref{S2_eq_resolventfree1-}).
\qed


\subsection{Position-dependent QW}

For the position-dependent QW $U$, we define $R(\kappa )= (U-e^{i\kappa} )^{-1} $ for $\kappa \in {\bf C} \setminus {\bf R} $.
The limiting absorption principle for $ R(\kappa )$ holds as follows.

\begin{lemma}
Following assertions hold.
\begin{enumerate}
\item For $ \theta \in J_{0} \setminus J_{0,\mathcal{T}} $ and $f\in \mathcal{B} ({\bf Z} )$, there exist weak-$*$ limits $R (\theta \pm i0 )f := \lim _{\epsilon \downarrow 0 } R ( \theta -i\log (1\mp \epsilon ))f$ in the sense
$$
\lim _{\epsilon \downarrow 0} (R ( \theta -i\log (1\mp \epsilon ))f,g) = (R(\theta \pm i0)f,g), \quad g\in \mathcal{B} ({\bf Z} ) .
$$

\item $R ( \theta \pm i0 )f$ for $f\in \mathcal{B} ({\bf Z} )$ satisfy $\| R (\theta \pm i0 )f\| _{\mathcal{B}^* ({\bf Z} ) } \leq c\| f\| _{\mathcal{B} ({\bf Z} )} $ for a constant $c>0$ which is independent of $\theta$ if $\theta$ varies over a compact interval in $J_{0} \setminus J_{0,\mathcal{T}} $.
\item The mappings $J_{0} \setminus J_{0,\mathcal{T}} \ni \theta \mapsto (R (\theta \pm i0 )f,g) $ are continuous for $f,g\in \mathcal{B} ({\bf Z} )$.
\end{enumerate}
\label{S3_lem_LAP}
\end{lemma}

Proof.
See Theorem 4.3 in \cite{Mo}.
\qed

\medskip

In view of the well-known resolvent equation
$$
R(\kappa )= R_0 (\kappa ) \left( 1-VR(\kappa ) \right) , \quad V=U-U_0 , \quad \kappa \in {\bf C} \setminus {\bf R} ,
$$
Lemmas \ref{S3_lem_asymptopticfree} and \ref{S3_lem_LAP} imply the following asymptotic behavior.

\begin{lemma}
Let us define the distorted Fourier transformation $ \mathcal{F}_{\pm} (\theta )$ by
\begin{equation}
\mathcal{F} _{\pm} (\theta )=( \mathcal{F} _{\pm ,1} (\theta ), \mathcal{F} _{\pm,2} (\theta )), \quad
\mathcal{F}_{\pm,j} (\theta )= \mathcal{F}_{0,j} (\theta )(1-VR(\theta \pm i0 )).
\label{S3_eq_fpm}
\end{equation}
Taking a sequence $f\in \mathcal{B} ({\bf Z} )$, we have 
\begin{gather}
( R (\theta + i0 )f)(x) \simeq  \sqrt{2\pi} e^{-i\theta} e^{ \pm i \theta x} (\mathcal{F}_{\pm ,1} (\theta )f) (\pm\theta ), \quad x\to \pm  \infty , \label{S3_eq_asymptotic1++}   \\
( R (\theta - i0 )f)(x) \simeq - \sqrt{2\pi} e^{-i\theta} e^{\mp i \theta x} (\mathcal{F}_{\pm,1} (\theta )f) (\mp \theta ), \quad x\to  \pm \infty , \label{S3_eq_asymptotic1+-}   
\end{gather}
for $ \theta \in J_{0,1} \setminus J_{0,\mathcal{T}}$, and
\begin{gather}
( R (\theta + i0 )f )(x)\simeq  \sqrt{2\pi} e^{-i\theta} e^{ \pm i \theta x} (\mathcal{F}_{\pm,2} (\theta )f) (\pm \theta ) , \quad x\to \pm \infty , \label{S3_eq_asymptotic2++}  \\
  ( R (\theta - i0 )f )(x)\simeq - \sqrt{2\pi} e^{-i\theta} e^{ \mp i \theta x} (\mathcal{F}_{\pm,2} (\theta )f) (\mp \theta ) , \quad x\to \pm \infty , \label{S3_eq_asymptotic2+-} 
\end{gather}
for $ \theta \in J_{0,2} \setminus J_{0,\mathcal{T}} $.
\label{S3_lem_asymptotic}
\end{lemma}

\textit{Remark.}
When $U-U_0$ is a sufficiently small perturbation, the perturbed Green function exists.
The perturbed Green function is the kernel function of $R(\kappa )$.
For our purpose in this paper, it is sufficient to use the resolvent equation and the distorted Fourier transformation.
The radiation condition for the equation $(U-e^{i\theta} )u=f$ can be determined by the asymptotic behavior of $G_0 (x,\theta \pm i0 )$ as $|x| \to \infty $.  
We study the asymptotic behavior of the Green function and its application to the radiation condition for a multi-dimensional QW in the forthcoming paper \cite{KKMS}.

\medskip

The adjoint operator $\mathcal{F}_{\pm} (\theta )^* $ of the distorted Fourier transformation characterizes the generalized eigenfunction of $U$.

\begin{lemma}
Let $\theta \in J_0 \setminus J_{0,\mathcal{T}} $.
Then we have $\mathcal{F}_{\pm} (\theta )^* {\bf h} (\theta )=\{ u\in \mathcal{B}^* ({\bf Z} ) \ ; \ (U -e^{i\theta} )u=0\} $.
\label{S3_lem_generalizedef}
\end{lemma}

Proof.
See Theorem 5.6 in \cite{Mo}.
\qed

\medskip

The generalized eigenfunction $ \mathcal{F}_+ (\theta )^* \phi$ derives the scattered wave as follows.
In view of the equality
$$
R(\theta +i0 ) ^* =-e^{i\theta} U R(\theta -i0),
$$
we have
$$
\mathcal{F}_+ (\theta )^* \phi = \mathcal{F}_0 (\theta )^* \phi +e^{2i \theta} R (\theta -i0) V^* \mathcal{F}_0 (\theta )^* \phi +e^{i\theta} V^* \mathcal{F}_0 (\theta )^* \phi .
$$
The third term on the right-hand side is negligible at infinity due to the assumption (A-1).

The S-matrix $\widehat{\Sigma} (\theta )$ for $ \theta \in J _0 \setminus J_{0,\mathcal{T}} $ is given by 
\begin{equation}
\widehat{\Sigma} (\theta )= 1-2\pi e^{i\theta} A(\theta ), \quad A(\theta )= \mathcal{F}_- (\theta ) V^* \mathcal{F}_0 (\theta )^* ,
\label{S3_eq_smatrix}
\end{equation}
and unitary on ${\bf h} (\theta )$.
For the proof, see Theorem 5.3 in \cite{Mo}.
Due to the definition of $\mathcal{F}_- (\theta )$ and the formula (\ref{S3_eq_generalef000}), we put
\begin{gather*}
\widehat{\Sigma}_j (\theta )=\left\{
\begin{split}
 1-2\pi e^{i\theta} \mathcal{F}_{-,j} (\theta ) V^* \mathcal{F}_0 (\theta )^* &, \quad \theta \in J_{0,j} , \\
1 &, \quad   \theta \not\in J_{0,j} ,
\end{split}
\right.
\end{gather*}
for $j=1,2$.
Then we have $\widehat{\Sigma} (\theta )=(\widehat{\Sigma}_1 (\theta ) , \widehat{\Sigma}_2 (\theta ))$, and it follows that each $ \widehat{\Sigma}_j (\theta )$ is a $2\times 2$ unitary matrix for $\theta\in J_{0,j} \setminus J_{0,\mathcal{T}} $.
In the following, we denote the S-matrix by 
$$
\widehat{\Sigma}_j (\theta )= \left[ \begin{array}{cc} \tau_j (\theta ) & \widetilde{\rho}_j (\theta ) \\ \rho_j (\theta ) & \widetilde{\tau}_j (\theta ) \end{array} \right] \in \mathrm{U} (2), \quad j=1,2.
$$

Now the asymptotic behavior of $ \mathcal{F}_+ (\theta ) ^* \phi $ comes from Lemma \ref{S3_lem_asymptotic} as follows.
The S-matrix naturally appears in the asymptotic behavior at infinity.

\begin{lemma}
Let $ \phi\in {\bf h} (\theta ) $.
We have 
\begin{gather}
( \mathcal{F}_+ (\theta )^* \phi )(x) \simeq \frac{1}{\sqrt{2\pi}} \left[ \begin{array}{c} \phi_j (\theta ) e^{i\theta x} \\ (\rho_j (\theta ) \phi_j (\theta )+ \widetilde{\tau}_j (\theta ) \phi_j (-\theta )) e^{-i\theta x}\end{array} \right] , \quad x\to \infty , \label{S3_eq_efasymptotic11}  \\
( \mathcal{F}_+ (\theta )^* \phi )(x) \simeq \frac{1}{\sqrt{2\pi}} \left[ \begin{array}{c} (\tau_j (\theta ) \phi_j (\theta ) + \widetilde{\rho}_j (\theta )\phi_j (-\theta))e^{i\theta x}  \\ \phi_j (- \theta ) e^{-i\theta x} \end{array} \right] , \quad x\to - \infty , \label{S3_eq_efasymptotic12}
\end{gather}
for $\theta \in J_{0,j} \setminus J_{0,\mathcal{T}} $. 
\label{S3_lem_efasymptotic}
\end{lemma}

In order to determine $\widehat{\Sigma} (\theta )$, it is sufficient to consider the cases $\phi_j (\theta )=\sqrt{2\pi}$, $\phi_j (-\theta )=0$, and $\phi_j (\theta )=0$, $\phi_j (-\theta )=\sqrt{2\pi}$.
Namely, we have 
\begin{gather}
( \mathcal{F}_+ (\theta )^* \phi )(x) \simeq  \left[ \begin{array}{c}  e^{i\theta x} \\ \rho_j (\theta )  e^{-i\theta x}\end{array} \right] , \quad x\to \infty , \label{S3_eq_asymptotic001} \\
( \mathcal{F}_+ (\theta )^* \phi )(x) \simeq  \left[ \begin{array}{c} \tau_j (\theta ) e^{i\theta x}  \\ 0 \end{array} \right] , \quad x\to - \infty , \label{S3_eq_asymptotic002}
\end{gather}
for $ \phi_j (\theta )=\sqrt{2\pi} $, $\phi_j (-\theta )=0$, and
\begin{gather}
( \mathcal{F}_+ (\theta )^* \phi )(x) \simeq  \left[ \begin{array}{c}  0 \\ \widetilde{\tau}_j (\theta )  e^{-i\theta x} \end{array} \right] , \quad x\to \infty ,\label{S3_eq_asymptotic003} \\
( \mathcal{F}_+ (\theta )^* \phi )(x) \simeq  \left[ \begin{array}{c} \widetilde{\rho}_j (\theta ) e^{i\theta x}  \\ e^{-i\theta x} \end{array} \right] , \quad x\to - \infty , \label{S3_eq_asymptotic004}
\end{gather}
for $ \phi_j (\theta )=0$, $\phi_j (-\theta )=\sqrt{2\pi}$.

In the following, we drop the suffix $j$ in every component of $ \widehat{\Sigma} _j (\theta )$ with respect to $ \theta \in J_{0,j} \setminus J_{0,\mathcal{T}} $ for the sake of simplicity of the notation.
$ \phi_L (\theta ) $ and $ \phi_R (\theta )$ denote $ \phi_j (\theta )$ and $ \phi_j (-\theta )$ for $ \theta \in J_{0,j} \setminus J_{0,\mathcal{T}} $, respectively. 
The S-matrix $\widehat{\Sigma} (\theta )$ for $ \theta \in J_{0,j} \setminus J_{0,\mathcal{T}} $ is simply denoted by 
$$
\left[ \begin{array}{cc} \tau (\theta ) & \widetilde{\rho} (\theta ) \\ \rho (\theta ) & \widetilde{\tau} (\theta ) \end{array} \right] \in \mathrm{U} (2) .
$$


\section{Combinatorial construction of scattering matrix}

\subsection{Finite rank perturbation}

In the following, we assume that the position-dependent QW $ U$ is a finite rank perturbation of the free QW $U_0 =S$.
We put $ \Gamma = \{ 0,1,\ldots ,n \} $ for a positive integer $n$, and 
\begin{gather}
C(x)= \left\{
\begin{split}
\left[ \begin{array}{cc} a(x) & b(x) \\ c(x) & d(x) \end{array} \right] ,&  \quad x\in \Gamma , \\
\left[ \begin{array}{cc} 1 & 0 \\ 0 & 1 \end{array} \right] ,&  \quad x\not\in \Gamma .
\end{split}
\right. 
\label{S4_eq_Cfinite}
\end{gather}
Due to the assumption (A-2), we have $a(x) \not= 0$ ($\Leftrightarrow d(x) \not= 0 $) for every $x\in \Gamma$.
Recall the rule of notation which has been introduced at the end of previous section.
Since $U-U_0$ is an operator of finite rank, the asymptotic behaviors (\ref{S3_eq_asymptotic001})-(\ref{S3_eq_asymptotic002}) and (\ref{S3_eq_asymptotic003})-(\ref{S3_eq_asymptotic004}) become exactly
\begin{gather}
( \mathcal{F}_+ (\theta )^* \phi )(x)= \left\{
\begin{split}
 \left[ \begin{array}{c} e^{i\theta x} \\ \rho (\theta ) e^{-i\theta x} \end{array} \right] , & \quad x\geq n+1 , \\
 \left[ \begin{array}{c} \tau (\theta ) e^{i\theta x} \\ 0 \end{array} \right] , & \quad x\leq -1 ,
\end{split}
\right.
\label{S3_eq_asymptotic001compact}
\end{gather}
for $ \phi_L (\theta )=\sqrt{2\pi}$, $\phi_R (\theta )=0$, and
\begin{gather}
( \mathcal{F}_+ (\theta )^* \phi )(x)= \left\{
\begin{split}
 \left[ \begin{array}{c} 0 \\ \widetilde{\tau} (\theta ) e^{-i\theta x} \end{array} \right] , & \quad x\geq n+1 , \\
 \left[ \begin{array}{c} \widetilde{\rho} (\theta ) e^{i\theta x} \\ e^{-i\theta x} \end{array} \right] , & \quad x\leq -1 ,
\end{split}
\right.
\label{S3_eq_asymptotic002compact}
\end{gather}
for $ \phi_L (\theta )=0$, $\phi_R (\theta )=\sqrt{2\pi}$.

In order to introduce a combinatorial construction of the S-matrix, let us relate the generalized eigenfunction of $U$ and the dynamics of the QW. 
Let $ \chi : \mathcal{B}^* ({\bf Z} ) \to \ell^2 (\Gamma ; {\bf C}^2 ) $ be defined by $ (\chi u)(x)=u(x)$ for $x\in \Gamma$ and $u\in \mathcal{B}^* ({\bf Z} )$.
The operator $\chi^* : \ell^2 (\Gamma ; {\bf C}^2 ) \to \mathcal{B}^* ({\bf Z} )$ is defined by 
\begin{gather*}
(\chi^* \psi )(x)= \left\{
\begin{split}
\psi (x) , & \quad x\in \Gamma , \\
0 , & \quad x\not\in \Gamma ,
\end{split}
\right.
\quad \psi \in \ell^2 (\Gamma ; {\bf C}^2 ).
\end{gather*}
Here the operator $\chi^* \in {\bf B} (\ell^2 (\Gamma ;{\bf C}^2 ); \ell^2 ({\bf Z};{\bf C}^2 ))$ is the adjoint operator of $\chi \in {\bf B} (\ell^2 ({\bf Z};{\bf C}^2) ;\ell^2 (\Gamma ;{\bf C}^2 )) $.
Note that $ \chi \chi^* $ is the identity on $\ell^2 (\Gamma ; {\bf C}^2 )$, and $ \chi ^* \chi $ is the projection onto the subspace $\{ u\in \mathcal{B}^* ({\bf Z} ) \ ; \ \mathrm{supp} u \subset \Gamma \} $.
Now we define the $2(n+1) \times 2(n+1)$ submatrix $E_n$ of $U$ by $E_n = \chi U \chi^* $.
Precisely, identifying a vector $\psi\in \ell^2 (\Gamma ; {\bf C}^2 )$ with $[\psi_R (0), \psi_L (0), \psi_R ( 1), \psi_L (1), \ldots , \psi_R (n), \psi_L (n)]^{\mathsf{T}} \in {\bf C} ^{2(n+1)}$, we have 
$$
E_n = \left[ \begin{array}{ccccc}
0 & P(1) & & & \\
Q(0) & 0 & P(2) & & \\
 & Q(1) & 0 &  & \\
& & \ddots & \ddots & \\
 & & & 0  & P(n) \\
 & &  & Q(n-1) & 0
\end{array} \right] ,
$$
where
$$
P(x)= \left[ \begin{array}{cc} 0 & 0 \\ b(x) & a(x) \end{array} \right] , \quad Q(x)= \left[ \begin{array}{cc} d(x) & c(x) \\ 0 & 0 \end{array} \right] ,
$$
for $x\in \Gamma $.

\begin{lemma}
Eigenvalues of $E_n$ lie in $\{ \lambda \in {\bf C} \ ; \ |\lambda | <1 \} $.

\label{S4_lem_evEn}
\end{lemma}

Proof.
Let $\psi \in \ell^2 (\Gamma ; {\bf C}^2 )$ be an eigenvector associated with the eigenvalue $\lambda$ of $E_n$.
Then we have 
\begin{equation}
|\lambda| ^2 \| \psi \| ^2_{\ell^2 (\Gamma ;{\bf C}^2 )} = \| E_n \psi \|^2 _{\ell^2 (\Gamma ; {\bf C}^2 )}   \leq ( U\chi^* \psi , U\chi^* \psi    ) _{\ell^2 ({\bf Z};{\bf C}^2 )} = \| \psi \|^2 _{\ell^2 (\Gamma ;{\bf C}^2 )} .
\label{S4_eq_evEn11}
\end{equation}
Now we suppose $|\lambda |=1$.
Then the equality in (\ref{S4_eq_evEn11}) holds.
Due to the definition of $\chi $ and $\chi^* $, this implies $\chi^* \chi U\chi^* \psi = U\chi^* \psi $ on ${\bf Z}$ so that $U\chi^* \psi = \lambda \chi^* \psi $.
This means that $\chi^* \psi $ is an eigenfunction of $U$ associated with the eigenvalue $\lambda$ with a finite support. 
By the assumption of $C(x)$, there is no eigenfunction of $U$ with a finite support. 
This is a contradiction.
We obtain $|\lambda|<1$.
\qed

\medskip

Under the setting of this paper, we can prove the convergence of the series $1+e^{-i\theta} E_n + e^{-2i\theta} E_n^2 + \cdots $ by using the Jordan canonical form of the matrix $E_n $.
Let $\lambda_1 , \ldots , \lambda_r $, $1\leq r\leq 2(n+1)$, be eigenvalues of $E_n$ with algebraic multiplicities $k_j$, $j=1,\ldots ,r$.
There exists a $2(n+1)\times 2(n+1)$ regular matrix $G_n$ such that
$$
G^{-1} _n E_n G_n = J(\lambda_1 ,k_1 )\oplus \cdots \oplus J(\lambda_r ,k_r ) ,
$$ 
where $J(\lambda_j ,k_j )$ is the Jordan block associated with the eigenvalue $\lambda_j$.

\begin{lemma}

The Neumann series $1+e^{-i\theta} E_n + e^{-2i\theta} E_n^2 + \cdots $ converges.
\label{S4_lem_diagEn}
\end{lemma}

Proof.
This lemma is a consequence of the Jordan canonical form of $E_n$ and Lemma \ref{S4_lem_evEn}.
\qed

\medskip

\textit{Remark.}
When $C(x)$ depends on the position $x\in \Gamma$, it is difficult to know the eigenvalues of $E_n$ in detail.
We can show the following facts.
\begin{itemize}
\item $E_n$ has $0$ as an eigenvalue. 
The dimension of the corresponding eigenspace is $2$ or higher.

\item The dimension of the eigenspace associated with a non-zero eigenvalue of $E_n$ is $1$. 
\end{itemize}

\medskip

The following fact is a special case of Theorem 3.1 in \cite{HiSe}. 
We derive a shortcut of the proof under our settings.

\begin{prop}
Let $\Psi_0 \in \mathcal{B}^* ({\bf Z} )$ be given by
\begin{gather*}
\Psi_0 (x)= \left\{
\begin{split} 
\alpha_L e^{i\theta x} \left[ \begin{array}{c} 1 \\ 0 \end{array} \right] , & \quad x\geq n+1, \\
\alpha_R e^{-i\theta x}  \left[ \begin{array}{c} 0 \\ 1 \end{array} \right]  , & \quad x\leq -1, \\
 \left[ \begin{array}{c} 0 \\ 0 \end{array} \right]  , & \quad \text{otherwise}, 
\end{split}
\right.
\end{gather*} 
for $\theta \in J_0 \setminus J_{0,\mathcal{T}} $, and for any constants $ \alpha_L , \alpha_R \in {\bf C}$.
We define $\Psi_t$ for every positive integer $t$ by $\Psi_t = U \Psi_{t-1} $.
Then there exists the limit 
$$
\Psi _{\infty} (x):= \lim _{t\to \infty} e^{-i \theta t}  \Psi_t (x) , \quad x\in {\bf Z} ,
$$
and $\Psi _{\infty} \in \mathcal{B}^* ({\bf Z} )$ satisfies 
\begin{equation}
U\Psi _{\infty} = e^{i\theta} \Psi_{\infty} \quad \text{on} \quad {\bf Z} .
\label{S4_eq_evefconstruction}
\end{equation}
In particular, we have $ \Psi _{\infty} = \mathcal{F}_+ (\theta )^* \phi $ for $\phi_L (\theta )=\sqrt{2\pi} \alpha_L  $ and $\phi_R (\theta )=\sqrt{2\pi} \alpha_R  $.

\label{S4_prop_HiSe}
\end{prop}

Proof.
The initial state $\Psi_0$ represents the flow incoming from infinity.
The incoming flow is split into two parts at $x=0 $ and $ n$ by the time evolution.
One is reflected and another one is transmitted.
Once the flow comes out of $\Gamma$, it goes to infinity.
In view of this dynamics, we decompose $\Psi_t$ as 
$$
\Psi_t = \chi^* \chi \Psi_t + (1-\chi^* \chi )(\Psi_t -e^{it\theta} \Psi_0 )+e^{it\theta} (1-\chi^* \chi )\Psi_0 .
$$
The first term on the right-hand side is the state in $\Gamma$ at time $t$.
The second term and the third term on the right-hand side are the outgoing flow and the incoming flow, respectively.

We put $\phi_t = \chi \Psi_t$.
Then we have 
\begin{equation}
\phi_0 =0, \quad \phi_{t+1} = E_n \phi_t + e^{i\theta t} \chi U \Psi_0 , \quad t\geq 0.
\label{S4_eq_constef11}
\end{equation}
Here we have used the relation $\chi U(1-\chi^* \chi ) U^t \Psi_0 = e^{i\theta t} \chi  U \Psi_0 $.
This recurrence relation implies 
$$
\phi_{t} = e^{i\theta ( t-1)} \sum _{m=0}^{t-1} e^{-i\theta m} E_n^m \chi U \Psi_0 , \quad t\geq 1 .
$$
In view of Lemmas \ref{S4_lem_evEn} and \ref{S4_lem_diagEn}, the limit 
$$
\phi _{\infty} :=  \lim _{t\to \infty} e^{-i\theta t} \phi_t = e^{-i\theta} (1-e^{-i\theta} E_n )^{-1} \chi U \Psi_0 
$$
 exists.
This implies 
\begin{equation}
E_n \phi_{\infty} + \chi U \Psi_0 = e^{i\theta} \phi _{\infty} .
\label{S4_eq_constef12}
\end{equation}

Let us turn to the outgoing flow.
We can see 
\begin{gather*}
\begin{split}
&(1-\chi^* \chi ) (\Psi_1 - e^{it\theta} \Psi_0 )=0, \\
&(1-\chi^* \chi ) (\Psi_t - e^{it\theta} \Psi_0 )= (1-\chi^* \chi ) \sum _{m=1}^{t-1} U^m f _{t-m} , \quad t\geq 2,
\end{split}
\end{gather*}
where the source $f_{t-m}$ is defined by $ f_{t-m} = (\delta _{0}+\delta_{n} ) \chi^* \phi _{t-m} $ for $\delta_y=\{ \delta_{xy} \} _{x\in {\bf Z}}$, $y\in {\bf Z} $.
For sufficiently large $t$, the value of $(1-\chi^* \chi ) (\Psi_t - e^{it\theta} \Psi_0 )$ at a point $x\in {\bf Z}\setminus \Gamma$ is given by $U^{\mu_x} f_{t-\mu_x} $ where $\mu_x =-x$ for $x\leq -1$ and $\mu_x = x-n$ for $x\geq n+1$.
The limit $\Psi_{out} := \lim_{t\to \infty} e^{-it\theta}  (1-\chi^* \chi ) (\Psi_t - e^{it\theta} \Psi_0 ) $ is determined by the limits
$$
\lim _{t\to \infty} e^{-it\theta} ( U^{\mu_x} f_{t-\mu_x} )(x) = e^{-i\mu_x \theta} (U f_{\infty} ) (x _{\pm} ) , \quad x\in {\bf Z} \setminus \Gamma ,
$$
where $f_{\infty} = (\delta_0 + \delta _n ) \chi^* \phi _{\infty}$, and $x_- =-1 $ for $x\leq -1 $ or $x_+ =n+1$ for $x\geq n+1$.
Thus we obtain 
\begin{equation}
\Psi_{out} (x)=e^{-i\mu_x \theta} (U f_{\infty} ) (x _{\pm} ) , \quad x\in {\bf Z} \setminus \Gamma .
\label{S4_eq_constef13}
\end{equation}

Finally, we prove the equation (\ref{S4_eq_evefconstruction}).
Note that 
$$
\Psi _{\infty} = \chi^* \phi _{\infty} + \Psi _{out} + \Psi _0 \quad  \text{on} \quad  {\bf Z}.
$$
From this decomposition, we have 
\begin{equation}
\chi U \Psi _{\infty} = E_n \phi _{\infty} + \chi U \Psi_0 + \chi U \Psi _{out}  = e^{i\theta} \phi_{\infty} ,
\label{S4_eq_constef14}
\end{equation}
due to (\ref{S4_eq_constef12}) and $\chi U \Psi_{out} =0$.
On the other hand, we have
\begin{gather}
\begin{split}
(1-\chi^* \chi ) U \Psi _{\infty} &= (1-\chi^* \chi )U (\chi^* \phi _{\infty} + \Psi _0 ) + U \Psi _{out} \\
&= (1-\chi^* \chi )\delta _{x_{\pm}} U f_{\infty} +e^{i\theta} \Psi _0  +e^{i \theta} \Psi _{out} -\delta _{x_{\pm}} Uf_{\infty} \\
&= e^{i\theta} (\Psi _{out} +\Psi _0 ) .
\end{split}
\label{S4_eq_constef15}
\end{gather}
Here we have used the relation $U \Psi _{out} = e^{i \theta} ( \Psi _{out} - \delta_{x_{\pm}} \Psi _{out} )$.
Plugging (\ref{S4_eq_constef14}) and (\ref{S4_eq_constef15}), we obtain the equation (\ref{S4_eq_evefconstruction}).
\qed

\medskip

Let us put $ \alpha _L = 1$ and $ \alpha_R =0$.
Then the generalized eigenfunction $\Psi _{\infty}$ satisfies the behavior (\ref{S3_eq_asymptotic001compact}).
The values $\Psi_{out} (-1)$ and $\Psi _{out} (n+1)$ determine $\tau (\theta )$ and $\rho  (\theta )$, respectively.
In the following, $A_{k,l}$ denotes the $(k,l)$-component of a matrix $A$.

\begin{theorem}
For $ \theta \in J_{0,j} \setminus J_{0,\mathcal{T}} $, we have 
\begin{gather*}
\begin{split}
& \tau  (\theta ) =a(0) (E_n^n (1-(e^{-i\theta} E_n )^2 )^{-1})_{2,2n+2} , \\
&\rho  (\theta )= e^{2in\theta } \left(  c(n) e^{i\theta} + d(n) e^{-2i\theta}   (E_n^2 (1-(e^{-i\theta} E_n )^2 )^{-1})_{2n+1,2n+2} \right) , \\
&\widetilde{\tau} (\theta )= d(n) ((E_n^n (1-(e^{-i\theta} E_n)^2))^{-1} ) _{2n+1,1} , \\
&\widetilde{\rho} (\theta )= a(0) e^{-2i\theta} (E_n^2 (1-(e^{-i\theta} E_n )^2 )^{-1} )_{2,1} +b(0)e^{i\theta} .
\end{split}
\end{gather*}
\label{S4_thm_smatrix}
\end{theorem}

Proof.
Suppose $ \alpha _L =1 $ and $ \alpha_R =0 $.
We have $( \Psi _{out} )_L (-1) = \tau  (\theta ) e^{-i\theta} $ and $ (\Psi _{out} )_R (n+1 )=\rho (\theta ) e^{-i\theta (n+1)}$ in view of (\ref{S3_eq_asymptotic001compact}).
We compute $ (\Psi _{out} )_L (-1)$ and $(\Psi _{out} )_R (n+1)$ by using (\ref{S4_eq_constef13}).

\begin{figure}[t]
\centering
\includegraphics[width=8cm, bb=0 0 656 391]{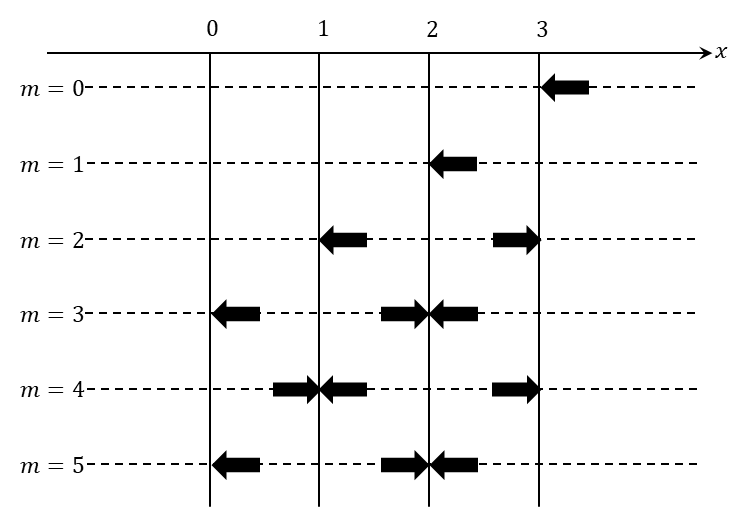}
\caption{The dynamics of $E_n^m \varphi $ for the case $n=3 $ implies the formulas of $( \phi _{\infty})_L (0)$ and $(\phi _{\infty} )_R (n)$. Leftwards arrows and rightwards arrows on every point represent $(E_n^m \phi )_L (x)$ and $(E_n^m \phi )_R (x)$, respectively.}
\label{S4_fig_dynamics}
\end{figure}

We have 
$$
\Psi _{out} (-1)= e^{-i\theta} \left[ \begin{array}{cc} a(0) ( \phi _{\infty} )_L (0) + b(0) (\phi _{\infty} )_R (0) \\ 0 \end{array} \right] .
$$
In view of $ \phi_{\infty} = e^{-i\theta} (1+e^{-i\theta} E_n + e^{-2i\theta} E_n^2 +\cdots ) \varphi $ where $\varphi = \chi U \Psi_0$, we compute $\phi_{\infty} (0)$ as follows (see also Figure \ref{S4_fig_dynamics}).
Note that 
$$
\varphi = \left[ \begin{array}{c} \varphi_R (0) \\ \varphi _L (0) \\ \vdots \\ \varphi_R (n) \\ \varphi _L (n) \end{array} \right] = \left[ \begin{array}{c} 0 \\ 0 \\ \vdots \\ 0 \\ e^{i\theta (n+1)} \end{array} \right]  .
$$

It follows $(\phi_{\infty} )_R (0)=0$ from $ \alpha_R =0$.
On the other hand, we have 
\begin{gather*}
\begin{split}
(\phi _{\infty} )_L (0) &= e^{-i \theta } e^{i\theta (n+1)} (e^{-i\theta n}  E_n^n + e^{-i\theta (n+2)} E_n^{n+2} +\cdots )_{2,2n+2} \\
&=(E_n^n (1- (e^{-i\theta} E_n )^2 )^{-1} ) _{2,2n+2} .
\end{split}
\end{gather*}
Then we have $ (\Psi _{out})_L(-1) = \tau  (\theta ) e^{-i\theta} = e^{-i \theta} a(0)  (E_n^n (1- (e^{-i\theta} E_n )^2 )^{-1} ) _{2,2n+2}$.
This equality implies the formula of $\tau (\theta )$.

Next we compute $\phi _{\infty} (n)$.
Note that 
$$
\Psi _{out} (n+1)= e^{-i\theta} \left[ \begin{array}{c} 0 \\ c(n) (\phi _{\infty} )_L (n)+d(n) (\phi _{\infty} )_R (n) \end{array} \right] .
$$
We have $ (\phi _{\infty} )_L (n)= e^{i\theta (n+1)} $ and 
\begin{gather*}
\begin{split}
(\phi _{\infty} )_R (n) &= e^{-i\theta} e^{i \theta (n+1)} (e^{-2i\theta} E_n^2 + e^{-4i \theta} E_n^4 +\cdots ) _{1,2n+2} \\
&= e^{i\theta (n-2)} ( E_n^2 (1-(e^{-i\theta} E_n )^2 )^{-1} )_{1,2n+2} .
\end{split}
\end{gather*}
It follows from this equality that $\Psi _{out} (n+1)= \rho (\theta  ) e^{-i\theta (n+1)} = e^{-i \theta} ( c(n) e^{i\theta (n+1)} + d(n) e^{ i\theta (n-2)}  ( E_n^2 (1-(e^{-i\theta} E_n )^2 )^{-1} )_{1,2n+2}) $.
Thus we obtain the formula of $ \rho (\theta ) $.

Taking $ \alpha_L =0$ and $ \alpha_R =1$ and repeating the similar argument, we also have formulas of $\widetilde{\tau} (\theta )$ and $ \widetilde{\rho} (\theta )$.
\qed


\subsection{Resonant-tunneling effect for double barrier model}
 
Let us consider the ``double-barrier" model : $C(x)$ is given by 
\begin{gather*}
C(x)= \left\{ \begin{split}
\left[ \begin{array}{cc} a_j & b_j \\ c_j & d_j \end{array} \right] &, \quad x = x_j, \\
\left[ \begin{array}{cc} 1 & 0 \\ 0  & 1 \end{array} \right] &, \quad x\not= x_j ,
\end{split}
\right. 
\end{gather*}
for $x_0 = 0$ and $ x_1 =n$ where $a_j \not= 0$ for $j=0,1$.
In this case, we can see the explicit formula of $\widehat{\Sigma} ( \theta )$ for $ \theta \in J_{0} \setminus J_{0,\mathcal{T}} $.

\begin{prop}
For $ \theta \in J_0 \setminus J_{0,\mathcal{T}} $, we have 
\begin{gather*}
\begin{split}
& \tau ( \theta )= \frac{a_0 a_1}{1-b_1 c_0 e^{-2i\theta n}} , \quad \rho (\theta )= \frac{c_0 \Delta_1 + c_1 e^{2i\theta n}}{1-b_1 c_0 e^{-2i\theta n}} , \\
& \widetilde{\tau} (\theta )= \frac{d_0 d_1}{1-b_1 c_0 e^{-2i\theta n}} , \quad \widetilde{\rho} (\theta )= \frac{b_1 \Delta_0 e^{-2i\theta n} +b_0}{1-b_1 c_0 e^{-2i\theta n}} ,
\end{split}
\end{gather*}
for $\Delta_j = a_j d_j - b_j c_j $.
\label{S4_prop_2barrier}
\end{prop}

Note that we can compute the generalized eigenfunction which satisfies (\ref{S3_eq_asymptotic001compact}) or (\ref{S3_eq_asymptotic002compact}).
Proposition \ref{S4_prop_2barrier} can be proven directly.

Now we consider the \textit{resonant-tunneling effect} as an analogue of the one dimensional Schr\"{o}dinger equation.
Resonant-tunneling effect means that an incident wave passes through a  barrier without loss of its energy.
In view of the scattering theory, this phenomenon can be derived by $\rho (\theta )=0 (\Leftrightarrow |\tau (\theta )|=1)$.

\begin{lemma}
If $ \rho (\theta )=0$ for some $ \theta \in J_0 \setminus J_{0,\mathcal{T}} $, there exist parameters $\alpha_j , \beta_j , \gamma_j \in {\bf T} $, $j=0,1$, and $p,q\in [0,1] $, $p\not= 0$, $p^2 +q^2 =1$, such that 
\begin{equation}
C(x_j)= e^{i\gamma_j /2} \left[ \begin{array}{cc} pe^{i(\alpha_j - \gamma_j /2)} & q e^{i(\beta_j -\gamma_j /2)} \\ -q e^{-i (\beta_j - \gamma_j /2)} & p e^{-i (\alpha_j - \gamma_j /2)} \end{array} \right] ,
\label{S4_eq_barrier}
\end{equation}
for $ j=0,1$.
$\rho (\theta )=0$ if and only if $ \widetilde{\rho} (\theta )=0 $.

\label{S4_lem_RTEF}
\end{lemma}

Proof.
Suppose $ \rho (\theta )=0 $.
We have $ c_0 \Delta_1 + c_1 e^{2i\theta n} =0 $.
In view of $|\Delta_1 |=1$, we have $|c_0 |=| c_1 |$.
It follows from the unitarity of $ C(x)$ that $|a_0 |=|a_1 |$, $|b_0 |= |b_1 | $, and $| d_0 |= |d_1 |$.
We obtain (\ref{S4_eq_barrier}).
Since $C(x)$ is unitary, the equivalence of $\widetilde{\rho} (\theta )=0$ and $\rho (\theta )=0$ is a consequence.
\qed

\medskip

The resonant-tunneling effect for the QW $U$ is derived as follows.

\begin{theorem}
Suppose that the matrix $ C(x_j) $ is given by (\ref{S4_eq_barrier}) for $j=0,1$.
If $q\not= 0$, we have $ \rho (\theta )=0 $ for 
$$
\theta = \frac{\beta_1 - \beta_0 + \gamma_0 +\pi}{2n} +\frac{\pi m}{n} ,
$$
where $m=0,1,\ldots ,2n-1$.
If $q=0$, we have $\rho (\theta )=0$ for any $\theta$.

\label{S4_thm_RTEF}
\end{theorem}

Proof.
We prove for the case $q\not= 0$.
In view of the formula (\ref{S4_eq_barrier}), $\rho (\theta )=0$ implies 
$$
0= c_0 \Delta_1 + c_1 e^{2i\theta n} =-qe^{i(\gamma_0 - \beta_0 )} e^{i\gamma_1} -q e^{i(\gamma_1 -\beta_1 )} e^{2i\theta n}.
$$
Then we have $ e^{2i\theta n} = e^{i(\beta_1 - \beta_0 + \gamma_0 +\pi )} $.
We have proven the theorem.
\qed

\medskip

As a direct consequence of Theorem \ref{S4_thm_RTEF}, we see that an inverse scattering problem can be solved.
Namely, we can determine $x_1 =n $ by the number of $ \theta $ such that $ \rho ( \theta )=0 $ for $q\not= 0 $.

\begin{cor}
Suppose that the matrix $ C(x_j) $ is given by (\ref{S4_eq_barrier}) for $j=0,1$, and $q\not= 0$.
If there exist $2n$ number of $\theta\in J_0 \setminus J_{0,\mathcal{T}} $ such that $ \rho (\theta )=0 $, we have $x_1 =n$.

\label{S4_cor_IRTEF}
\end{cor}


\subsection{Incident wave always passes through a barrier}

As has been seen in the previous subsection, $\rho (\theta )$ may vanish for some $\theta \in J_{0} \setminus J_{0,\mathcal{T}} $.
At the end of this paper, let us show that $\tau (\theta )$ does not vanish for all $\theta \in J_{0} \setminus J_{0,\mathcal{T}} $ under the assumption (A-2).

\begin{theorem}
Suppose that $C(x)$ is given by (\ref{S4_eq_Cfinite}).
The transmission coefficient $\tau (\theta )$ does not vanish for all $\theta \in J_0 \setminus J_{0,\mathcal{T}} $.
\label{S4_thm_alwayspasses}
\end{theorem}

Proof.
Suppose $\tau (\theta )=0$ for some $\theta \in J_0 \setminus J_{0,\mathcal{T}} $.
In view of (\ref{S3_eq_asymptotic001compact}), we have $(\mathcal{F}_+ (\theta )^* \phi )(x) =0$ for $x\leq -1$.
Since $u : = \mathcal{F}_+ (\theta )^* \phi $ is a generalized eigenfunction of $U$, we have 
$$
\left[ \begin{array}{cc} e^{i\theta} & 0 \\ c(x) & d(x) \end{array} \right] \left[ \begin{array}{c} u_L (x) \\ u_R (x) \end{array} \right] -  \left[ \begin{array}{cc} a(x+1) & b(x+1) \\ 0 & e^{i\theta} \end{array} \right] \left[ \begin{array}{c} u_L (x+1) \\ u_R (x+1) \end{array} \right] =0.
$$
It follows 
\begin{equation}
\left[ \begin{array}{c} u_L (x+1) \\ u_R (x+1) \end{array} \right] = \left[ \begin{array}{cc} a(x+1) & b(x+1) \\ 0 & e^{i\theta} \end{array} \right] ^{-1} \left[ \begin{array}{cc} e^{i\theta} & 0 \\ c(x) & d(x) \end{array} \right]  \left[ \begin{array}{c} u_L (x) \\ u_R (x) \end{array} \right] ,
\label{S4_eq_EE1}
\end{equation}
for any $x\in {\bf Z} $.
Note that the inverse matrix on the right-hand side exists due to the assumption (A-2).
Since $u(x)=0$ for $x\leq -1 $, we have $ u(x)=0$ for any $x\in {\bf Z} $ by using the equation (\ref{S4_eq_EE1}).
This is a contradiction.
\qed

\medskip

If the assumption (A-2) does not hold, the incident wave is reflected at a point.
This is a trivial case where there exist complete reflection phenomena of quantum walkers.
For example, we take 
\begin{gather*}
C(x)= \left\{ \begin{split}
\left[ \begin{array}{cc} 0 & b \\ c & 0 \end{array} \right] &, \quad x = 0,n, \\
\left[ \begin{array}{cc} 1 & 0 \\ 0  & 1 \end{array} \right] &, \quad x\not= 0,n  ,
\end{split}
\right.   
\end{gather*}
 where $|b|=|c|=1$.
Then a quantum walker is reflected at $x=0$ and $n$.
Moreover, there  exists an eigenfunction $u$ of $U$ such that $\mathrm{supp} u \subset [0,n]$ (see \cite{MoSe}).
The matrix $ C(x) $ gives a model of non-penetrable barrier at $x=0$ and $n$.
The assumption (A-2) guarantees the penetrability condition of barriers given by $C(x) $.

\vspace*{0.5cm}

\noindent
{\bf Acknowledgement.}
The authors greatly appreciate a valuable comment by Dr. Kenta Higuchi. In particular, his comment is helpful for the correction of Lemma \ref{S4_lem_diagEn}. 
H. Morioka is supported by the JSPS Grant-in-aid for young scientists (B) No. 16K17630 and the JSPS Grant-in-aid for young scientists No. 20K14327. 
E. Segawa is supported by the JSPS Grant-in-Aid for Scientific Research (C) No. 19K03616 and Research Origin for Dressed Photon.


\appendix

\section{Complex contour integration}

Here we compute the integration
\begin{equation}
I(x,\kappa )= \int _{-\pi}^{\pi} \frac{e^{ix\eta}}{-\cos \eta +\frac{1}{p} \cos (\kappa -\gamma /2)} d\eta , \quad x\in {\bf Z} ,\quad \kappa \in {\bf C} \setminus {\bf R} ,
\label{App_integral}
\end{equation}
by using the residue theorem.
In order to use the complex contour integral, we extend the integrand to the complex variable
$$
F(z,x,\kappa )=\frac{e^{ixz}}{-\cos z+\frac{1}{p} \cos (\kappa - \gamma/2)} , \quad z\in {\bf T} _{{\bf C}} .
$$
The integrand $F(z,x,\kappa )$ has simple poles at $z= \pm \zeta (\kappa )$ where
\begin{equation}
 \zeta (\kappa )= \arccos \left( \frac{1}{p} \cos \left( \kappa -\frac{\gamma}{2} \right) \right). \label{S2_eq_MCK}
\end{equation}
Moreover, we have 
$$
{\mathop{{\rm Res}} _{z=\pm \zeta (\kappa )}} F(z,x,\kappa ) = \pm \frac{e^{\pm ix\zeta (\kappa )}}{\sin \zeta (\kappa )} .
$$

We assume $\mathrm{Im} \, \zeta (\kappa )> 0$.
In order to compute $ I(x,\kappa )$, we consider the complex contour integral
$$
I_{\mathcal{C}} (x,\kappa )= \int _{\mathcal{C}} F(z,x,\kappa )dz ,
$$
where $\mathcal{C} = \sum_{j=0}^3 \mathcal{C}_j $ with
\begin{gather*}
\begin{split}
&\mathcal{C}_0 = \{ w=t \ ; \ t:-\pi \to \pi \} , \quad \mathcal{C}_1 = \{ w=\pi +it \ ; \ t:0\to \rho \} , \\
&\mathcal{C}_2 = \{ w=t+i\rho \ ; \ t:\pi \to -\pi \} , \quad \mathcal{C}_3 = \{ w=-\pi +it \ ; \ t:\rho \to 0 \} ,
\end{split}
\end{gather*}
for sufficiently large $\rho >0$.
Due to the residue theorem, we have 
$$
I_{\mathcal{C}} (x,\kappa )= \frac{2\pi i e^{ix\zeta (\kappa )}}{\sin \zeta (\kappa )} .
$$
In view of the periodicity of the integrand, we can see 
$$
\int _{\mathcal{C}_1 } +\int _{\mathcal{C}_3} F(z,x,\kappa )dz =0.
$$
We also have
$$
\left| \int _{\mathcal{C}_2} F(z,x,\kappa )dz \right| \leq ce^{-\rho (x+1)},
$$
for a constant $c>0$.
This implies 
$$
\int _{\mathcal{C}_2} F(z,x,\kappa )dz \to 0,
$$
as $\rho \to \infty $ for $x\geq 0$.
Now we obtain 
\begin{equation}
I (x,\kappa )=\frac{2\pi i e^{ix\zeta (\kappa )}}{\sin \zeta (\kappa )} , \quad x\geq 0.
\label{App_eq_Ik+}
\end{equation}

Let us turn to the complex contour integral
$$
I_{\mathcal{C}'} (x,\kappa )=- \frac{2\pi i e^{-ix\zeta (\kappa )}}{\sin \zeta (\kappa )},
$$
for $\mathcal{C}'= \sum_{j=0}^3 \mathcal{C}'_j$ with
\begin{gather*}
\begin{split}
&\mathcal{C}'_0 = \{ w=t \ ; \ t:\pi \to - \pi \} , \quad \mathcal{C}'_1 = \{ w=-\pi -it \ ; \ t:0\to \rho \} , \\
&\mathcal{C}'_2 = \{ w=t-i\rho \ ; \ t:-\pi \to \pi \} , \quad \mathcal{C}'_3 = \{ w=\pi -it \ ; \ t:\rho \to 0 \} ,
\end{split}
\end{gather*}
for sufficiently large $ \rho >0 $.
By the similar way for the case $\mathcal{C}$, we can see
\begin{equation}
I (x,\kappa )=\frac{2\pi i e^{-ix\zeta (\kappa )}}{\sin \zeta (\kappa )} , \quad x\leq 0.
\label{App_eq_Ik-}
\end{equation}
Plugging (\ref{App_eq_Ik+}) and (\ref{App_eq_Ik-}), we obtain the following fact for $\mathrm{Im} \, \zeta (\kappa )>0$.
For the case $\mathrm{Im} \, \zeta (\kappa )<0$, the proof is similar.

\begin{lemma}
For $x\in {\bf Z} $ and $\kappa \in {\bf C} \setminus {\bf R} $, we have 
\begin{equation}
I(x,\kappa )=\pm \frac{2\pi i e^{\pm i|x| \zeta (\kappa )}}{\sin \zeta (\kappa )} \quad \text{for} \quad \pm\mathrm{Im} \, \zeta (\kappa )>0.
\end{equation}
\label{App_lem_contourintegral}
\end{lemma}

\end{document}